\documentclass[12pt]{article}

\usepackage{amsmath,amssymb,array,amsfonts}
\usepackage[dvipdfmx]{graphicx}
\usepackage{comment,color}
\usepackage{ascmac}
\usepackage{cite}

\newcommand{\mot}{\!\not \!}
\newcommand{\mmot}{\!\!\not \!\! }

\newcommand{\ba}{\begin{eqnarray}}
\newcommand{\ea}{\end{eqnarray}}

\allowdisplaybreaks

\def\ncm{\newcommand}
\def\M {{\rm M}}
\def\e {{\rm e}}

\def\H {{\rm H}}
\def\B {{\rm B}}
\def\SM{{\rm SM}}
\def\T {{\rm T}}

\def\rmd{{\rm d}}

\def\dis{\displaystyle}

\def\nt{\notag}

\ncm{\sls}[1]{{\ooalign{\hfil/\hfil\crcr$#1$}} }

\makeatletter
    
    \@addtoreset{equation}{section}
\makeatother

\begin{document}
\setlength{\baselineskip}{18pt}

\begin{titlepage}

\begin{flushright}
OCU-PHYS-477
\end{flushright}
\vspace{1.0cm}
\begin{center}
{\Large\bf 
Revisiting Electroweak Symmetry Breaking and \\
\vspace*{3mm}
Higgs Mass in Gauge-Higgs Unification
} 
\end{center}
\vspace{25mm}

\begin{center}
{\large
Yuki Adachi 
and 
Nobuhito Maru$^{*}$
}
\end{center}
\vspace{1cm}
\centerline{{\it
Department of Sciences, Matsue College of Technology,
Matsue 690-8518, Japan.}}

\centerline{{\it
$^{*}$
Department of Mathematics and Physics, Osaka City University, Osaka 558-8585, Japan.
}}
%
%
\vspace{2cm}
\centerline{\large\bf Abstract}
\vspace{0.5cm}
We propose a new model of 5D $SU(3) \otimes U(1)_X$ gauge-Higgs unification 
 with a successful electroweak symmetry breaking and a realistic Higgs boson mass.  
In our model, the representations of the fermions are very simple,  
 the $\bf 3, \bar{\bf 3}$ and $ \overline{\bf 15}$ representations of $SU(3)$ gauge group.
Employing the anti-periodic boundary conditions for $\overline{\bf 15}$ reduces massless exotic fermions 
 and simplifies the brane localized mass terms. 
We calculate the 1-loop Higgs potential in detail and 
 find that a realistic electroweak symmetry breaking and the observed Higgs mass are obtained. 


\end{titlepage}

\newpage
\section{Introduction}

Gauge-Higgs unification (GHU) \cite{GH,HIL} unifies the standard model (SM) gauge boson and Higgs boson 
 into the higher dimensional gauge fields. 
This scenario is one of the attractive ideas that solves the hierarchy problem without invoking supersymmetry, 
 since the Higgs boson mass and its potential are calculable due to the higher dimensional gauge symmetry \cite{HIL}.
These characteristic features have been studied and verified in models 
 with various types of compactification at one-loop level \cite{ABQ}
 and at the two-loop level \cite{MY}. 
The calculability of other physical observables such as $S$ and $T$ parameters \cite{LM}, 
 Higgs couplings to digluons, diphotons \cite{Maru}, muon $g-2$ and the EDM of neutron \cite{ALM} 
 have been also investigated. 
 The flavor physics which is a very nontrivial in GHU has been studied in \cite{flavorGHU}.

In five dimensional (5D) GHU, 
 Higgs potential at the tree level is forbidden by the gauge symmetry in higher dimensions,  
 but it is radiatively generated. 
Due to its characteristic features, 
 it is nontrivial to obtain a realistic electroweak symmetry breaking and the observed Higgs mass. 
In GHU, Higgs quartic coupling is provided by the gauge coupling squared and is 1-loop suppressed.
\footnote{
Note that top quark contribution to the Higgs quartic coupling is also given by the gauge coupling squared  
 due to the fact that the Yukawa couplings are proportional to the gauge coupling in this scenario. 
The contribution is crucial in order to realize the electroweak symmetry breaking. 
}
Therefore, Higgs mass squared is naively of order 1-loop factor times the compactification scale squared 
 $m_h^2 \sim \frac{1}{16\pi^2 R^2}$. 
Noting that W boson mass and the compactification scale are related by $M_W = c/R $ 
 in terms of a dimensionless parameter ``$c$" 
 which is determined from the potential minimum, 
 Higgs mass is too small if the parameter $c$ is an order of the unity \cite{SSS}. 
If we manage to realize a small parameter $c$ by potential minimization, 
 this allows the larger compactification scale and heavier Higgs mass. 
In order to obtain a small parameter $c$, 
 it is well-known that it has to be generated 
 by the contributions from different representations of the gauge group. 
 
It is troublesome to eliminate the massless exotic fermions. 
Embedding the SM fermions into the large representations, 
 there exist many massless exotic fermions 
 and they are ordinary made massive by introducing the brane mass terns 
 and extra brane localized fermions coupling to Dirac mass with the exotic fermions. 
Even for the $SU(2)$ SM doublets, 
 the number of massless doublets is duplicated in each generation 
 since massless doublets appear from the isospin up and down components. 
To eliminate the half of the massless doublets, 
 we also must introduce the brane mass terms and extra brane localized doublets 
 coupling to Dirac mass with the exotic massless doublets. 
Such brane mass terms complicate models and analysis. 
Therefore, it is desirable to construct a model 
 where the brane mass terms are as little as possible.  
 
In this paper, we propose a new model of 5D $SU(3) \otimes U(1)_X$ GHU 
 with a successful electroweak symmetry breaking and a realistic Higgs boson mass.  
In our model, the representations of the fermions are only two kinds, that is, 
 the $\bf 3,\overline {\bf 15}$ for the third generation quarks 
 and $\bf 3 \oplus \bar{\bf 3}$ representations of $SU(3)$ gauge group 
 for other SM fermions.\footnote{The reason why the representation of third generation is only different from other SM fermions 
 is to generate top yukawa coupling. 
In GHU, 
 an enhancement factor is required to obtain the top yukawa coupling 
 since yukawa coupling is provided by a gauge coupling and it gives the W boson mass 
 after the electroweak symmetry breaking.}
In our setup, top quark is not embedded into the $\overline{\bf 15}$ representations. 
In this case, we have no need to obtain massless zero mode from the bulk fermions 
 and we can impose the anti-periodic boundary conditions for $\overline{\bf 15}$. 
Therefore, we have no need to introduce the brane localized mass terms 
 since the lightest mode is necessarily massive. 
From the fundamental representations, 
 the massless exotic $SU(2)_L$ doublets are unavoidable 
 because these massless doublets appear from the up- and down-type sectors. 
 Namely, the number of the massless $SU(2)_L$ doublets are doubled. 
We have to introduce the brane mass terms for one linear combinations 
 of doublets to make them massive. 

We calculate the 1-loop Higgs potential and search a viable matter content 
 to realize a realistic electroweak symmetry breaking and the observed Higgs mass. 
In order to accomplish this, 
 it is found that a pair of additional $\overline{\bf 15}$ representations 
 other than the SM fermions should be included. 
We also study whether the top and bottom quark masses are reproduced. 
Note that  the masses of $SU(2)_L$ doublets are correlated through the mixing 
 between the $SU(2)_L$ doublets from the up- and down-type sectors
 \footnote{These mixings are crucial for the flavor violation in the context GHU, see \cite{flavorGHU}}. 
For the third generation quarks, 
 it is not a trivial issue since the mass difference between top and bottom is larger 
 than those of the first two generations.  

This paper is organized as follows. 
In section 2, we describe our model. 
In section 3, we calculate the mass spectrum of the various fields introduced in our model.   
The Higgs potential is calculated and analyzed in section 4. 
Summary is devoted to section 5. 
In appendix, the details of several representations are summarized. 

\section{A model}
We consider the $SU(3)\otimes U(1)_X$ gauge theory in five-dimensional flat space-time.
The fifth extra dimension is compactified on an orbifold $S^1/Z_2$ where the radius of $S^1$ is $R$.
Because the weak mixing angle $\theta_W$ in the $SU(3)$ is not consistent with the realistic one,
the correct value is effectively realized by the mixing between the $U(1)_X$ and neutral gauge bosons of $SU(3)$.

\begin{figure}
\begin{center}
\includegraphics{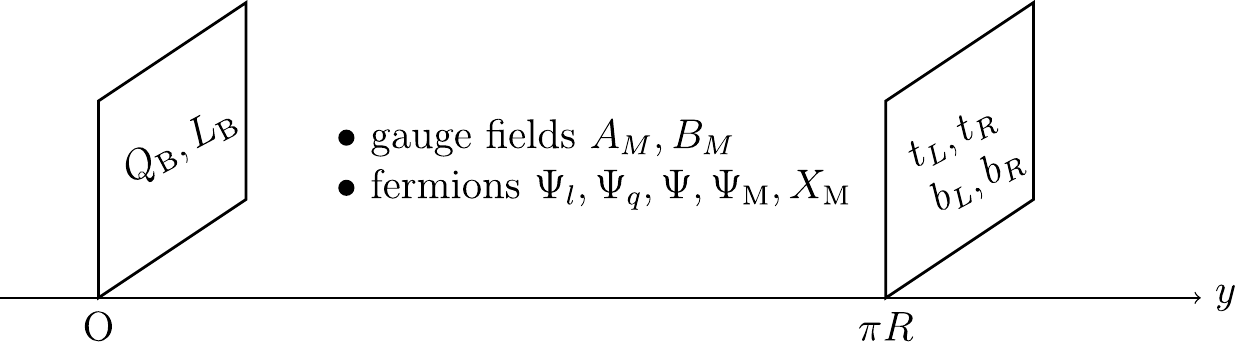}
\caption{Setup of the model.}
\label{fig:outline}
\end{center}
\end{figure}

The SM chiral fermions are introduced as follows: 
The top quark ($t$) and the bottom quark ($b$) quarks are brane-localized fermions localized at the $y=\pi R$ brane.
Other SM fermions are embedded in the bulk fermions $\Psi_l$ and $\Psi_q$.
They obtain a mass through the five-dimensional gauge interaction
 as ordinary way in the context of the gauge-Higgs unification scenario.
Since the $t$ and $b$ quarks cannot interact directly with Higgs boson ($A_y$),
 the extra bulk fermions $\Psi$ (referred to as messenger fermions) are necessary to connect them.
We also introduce a pair of fermions (referred to as mirror fermions)
$\Psi_\M$ and $X_\M$
to realize the realistic electroweak symmetry breaking. 
Such fermions may be a possible candidate of the dark matter
as pointed out in \cite{Maru:2017otg}.
The outline of this model is depicted in the Figure \ref{fig:outline}.
Such a strategy simplifies our model:
The top quark needs a large representation to reproduce the large top yukawa coupling 
as will be mentioned in the following sections.
In general, such a large representation includes the massless exotic fermions
but they are automatically removed from the low-energy effective theory by use of the anti-periodic boundary condition.
The other light fermions are embedded in the fundamental representations
to assign the suitable $U(1)_X$ charges.
Thus the extra brane fermions and brane mass terms are greatly reduced in our model.

Since the gauge sectors of our model has been discussed in detail \cite{Adachi:2016zdi},
we focus on the the fermion sector 
in the following subsections.

\subsection{The third generation quark}
In this subsection,
we discuss the $t$ and $b$ quarks.
As is mentioned in the previous paragraph,
 the third generation $t_R,b_R$ and 
 $\begin{pmatrix}
 t_L\\b_L
 \end{pmatrix}
 $
 are put on the $y=\pi R$ brane.
 The chirality is defined as the eigenvalues of the chiral projection operators
$L=(1-\gamma_5)/2, R=(1+\gamma_5)/2$.
As for the messenger fermion,
the  $\bf 3_0$ and $\overline{\bf 15}_{\bf -2/3}$ representations 
are introduced.
The subscripts stand for the $U(1)_X$ charges
in order to couple with the brane fermions.
These messenger fermions include the 
two quark doublets $Q_{3},Q_{15}$ and the two singlets $B,T$.

We impose the $Z_2$ symmetry and anti-periodic boundary condition
on the messenger fermions to leave the chiral fermions:
\begin{equation}
\Psi(y+2\pi R) = -\Psi(y), \quad 
\Psi(y)=-P\gamma_5 \Psi (-y) 
\end{equation}
where the matrix $P$ is defined as 
$P=\text{diag}(++-)$ for the $SU(3)$ fundamental representation.
Due to such an anti-periodic boundary condition, 
 they obtain a mass at least around $1/(2R)$, 
 so the exotic fermions are automatically removed from the low energy effective theory. 
Since the $Z_2$ parities at $y=\pi R$ is opposite to those of the $y=0$ because of the anti-periodicity,
the right-handed doublet and left-handed singlet in the messenger fermion can couple to the SM fermions ($t$ and $b$ quarks) at $y=\pi R$.
The Lagrangian of the third generation quarks becomes
\begin{align}
\mathcal L_\text{fermion}
\supset  &
\bar \psi ({\bf 3_0}) i\mmot D_{{\bf 3}}\psi ({\bf 3_0})
 +\bar \psi (\overline{\bf 15}_{\bf -2/3}) i\mmot D_{\overline{\bf 15}}\psi (\overline{\bf 15}_{\bf -2/3})
 \nonumber \\&
+\delta(y-\pi R)
\Big[
\bar Q_{L} i\partial_\mu\gamma^\mu Q_{L} 
+\bar t_R i\partial_\mu\gamma^\mu t_R
+\bar b_R i\partial_\mu\gamma^\mu b_R
\nonumber \\
&
+\frac{\epsilon_{L}}{\sqrt{\pi R}}Q_{L}
(\cos\theta Q_{15R}+\sin\theta Q_{3R})
+\frac{\epsilon_{tR}}{\sqrt{\pi R}} \bar T_L t_R
+\frac{\epsilon_{bR}}{\sqrt{\pi R}} \bar B_L b_R
+{\rm h.c.}
\Big].
\end{align}
The covariant derivative is $D^M=\partial^M+ig_5 A^M_a T^a+ig'_5 Q_X B^M~(M=0,1,2,3,4)$.
The $A^M$ and $B^M$ represent the gauge fields of $SU(3)$ and $U(1)_X$, respectively.
The $Q_X$ stands for the $U(1)_X$ charges.
The $g_5$ and $g_5'$ represents the five dimensional gauge couplings, respectively.
One can see from this Lagrangian that the SM chiral fermions $t_R,b_R$ and  $Q_L$ can interact 
 with the Higgs ($A_y$) through the messenger fermions.

\subsection{The first and second generations of quarks and leptons }
\label{appendix:lightergeneration}
We choose $\bf 3$ and $\bf{\bar 3}$ representations for the SM chiral fermions 
except for the $t$ and $b$ quarks.
They include the two doublets and two singlets as follows:
\begin{equation}
\Psi_q({\bf 3})\supset
\begin{pmatrix}
Q_{\bf 3}\\ d
\end{pmatrix}
,
\Psi_q({\bf{\bar 3}_{1/3}})\supset
\begin{pmatrix}
i\sigma_2 Q_{\bf{\bar 3}}\\ u
\end{pmatrix}
,
\Psi_l({\bf 3_{\bf -2/3}})\supset
\begin{pmatrix}
L_{\bf 3}\\ e
\end{pmatrix}
,
\Psi_l{(\bf{\bar 3}_{-1/3})}\supset
\begin{pmatrix}
i\sigma_2 L_{\bf{\bar 3}}\\ \nu
\end{pmatrix}
\end{equation}
We impose the $Z_2$ symmetry and periodic boundary conditions on the $\Psi_q$ 
and $\Psi_l$.
\begin{equation}
\Psi_{l,q}(+ y)=-P\gamma_5 \Psi_{l,q}(- y), \quad 
\Psi_{l,q}(y)=\Psi_{l,q}(y+2\pi R).
\end{equation}
Then the Lagrangian of the first two generation quarks is written as
\begin{align}
\mathcal L_\text{fermion}
\supset&\nt
\sum_{i=1}^2
\left\{
\bar\Psi^i_q({\bf 3_0}) \left[i\mot D_{\bf 3}+M_q^i\epsilon(y)\right] \Psi_q^i({\bf 3_0})
+\bar\Psi^i_q(\bar{\bf 3}_{\bf 1/3}) \left[i\mot D_{\bar{\bf 3}}+M_q^i\epsilon(y)\right] \Psi_q^i(\bar{\bf 3}_{\bf 1/3})
\right\}
\\&\nt
+
\sum_{i=1}^3
\left\{\bar\Psi^i_l({\bf 3}_{\bf -2/3}) \left[i\mot D_{\bf 3}+M^i_l\epsilon(y)\right] \Psi^i_l({\bf 3}_{\bf -2/3})
+\bar\Psi^i_l(\bar{\bf 3}_{\bf -1/3}) \left[i\mot D_{\bar{\bf 3}}+M^i_l\epsilon(y)\right] \Psi^i_l(\bar{\bf 3}_{\bf -1/3})
\right\}
\nonumber 
\\ \nt
&+ \delta(y)
\sum_{i=1}^2
 \left[
  \bar Q_\B^i i\partial_\mu\gamma^\mu Q^i_\B
  +\frac{\epsilon_q}{\sqrt{\pi R}}
  \bar Q^i_{\B}Q^i_{\rm H} +{\rm h.c.}
 \right]
 \\
&+ \delta(y)
\sum_{i=1}^3
 \left[
  \bar L_\B^i i\partial_\mu\gamma^\mu L^i_\B
  +\frac{\epsilon_l}{\sqrt{\pi R}}
  \bar L^i_{\B}L^i_{\rm H} +{\rm h.c.}
 \right]. 
\end{align}
The $Q_\B$ and $L_\B$ are the brane localized fermions which couples one of the duplicated doublets.
The $Q_\H$ and $L_\H$ are the linear combinations of doublets in the bulk fields.
The bulk mass term $M_q$ and $M_l$ give exponential suppressions like 
$e^{-M_{q} \pi R}$ or $e^{-M_{l} \pi R}$
to the yukawa couplings,
 the hierarchical fermion masses can be achieved by mild tuning of bulk mass parameters 
$M_{q}$  and $M_{l}$.

\subsection{Mirror fermions}
In our setup with the SM fermions and the messenger fermions, 
 we have to introduce further extra fermions 
 since the realistic electroweak symmetry breaking does not happen.
In this paper, a pair of the $\overline{\bf 15}$ representations, 
 which is referred to as mirror fermions, are introduced. 
They obey the $Z_2$ symmetry and periodic boundary conditions as follows:
\begin{equation}
\label{BC_mirror1}
\Psi_\M(+ y)=P\gamma_5 \Psi_\M(- y), \quad
\Psi_\M(y)=\Psi_\M(y+2\pi R).
\end{equation}
Similar boundary conditions are imposed on the $X_\M$ as
\begin{equation}
\label{BC_mirror2}
X_\M(+ y)=-P\gamma_5 X_\M(- y), \quad 
X_\M(y)=X_\M(y+2\pi R).
\end{equation}
Since these boundary conditions allow the massless chiral fermion in the zero mode,
 the bulk mass term is added to make them massive.
The Lagrangian of the mirror fermions is given by 
\begin{equation}
\mathcal L_\text{fermion}
\supset 
\bar\Psi_\M i\mot D \Psi_\M
+\bar X_\M i\mot D X_\M
+M
\left[\bar\Psi_\M X_\M +\bar X_\M\Psi_\M \right].
\end{equation}
As mentioned earlier, 
mirror fermions introduced in this subsection are interesting 
in that their lightest fermion might be a dark matter candidate.  
Such a pair of $\overline{\bf 15}$ representations are also natural 
from the viewpoint of the minimal dark matter scenario 
in the context of gauge-Higgs unification \cite{Maru:2017otg}. 


\section{Mass spectrum}
We discuss here the mass spectrum
 necessary for calculating the 1-loop effective potential for Higgs field.

\subsection{top and bottom sector}
The $\overline{\bf 15}$ representation of $SU(3)$ includes 
 the singlet $T$, doublet $Q_{15}$, triplet $\Sigma$, quartet $\Lambda$, 
and  quintet $\Delta$ of $SU(2)$.
Since the $\bf 3$ representation 
includes doublet $Q_3$ and singlet $B$,
there are two doublets 
$Q_3=\begin{pmatrix}
T_3\\B_3
\end{pmatrix}
$ and $Q_{15}=\begin{pmatrix}
T_{15}\\B_{15}
\end{pmatrix}$ in our model.
Adopting the vector notation 
$\vec T=(T,T_3,T_{15},\Sigma_t,\Lambda_t,\Delta_t)$
and
$\vec B=(B,B_3,B_{15},\Sigma_b,\Lambda_b,\Delta_b)$, 
 the quadratic part of Lagrangian for the top and bottom quarks are
\begin{align}
\mathcal L_\text{fermion}
\supset &
\bar \psi ({\bf 3_0}) i\mmot D_{{\bf 3}}\psi ({\bf 3_0})
 +\bar \psi (\overline{\bf 15}_{\bf -2/3}) i\mmot D_{\overline{\bf 15}}\psi (\overline{\bf 15}_{\bf -2/3}),
 \\
\supset&
\vec {\bar T}^\T i\partial_M\Gamma^M \vec T
+\vec {\bar B}^\T i\partial_M\Gamma^M \vec B
+
M_W
\left[\vec {\bar T}^\T M_t\gamma^5 \vec T+\vec {\bar B}^\T M_b\gamma^5 \vec B \right]
\end{align}
where the covariant derivatives are given by
\begin{equation}
D_{{\bf 3}M} =  \partial_M  - ig_5 A_M,
 \quad
D_{\overline{\bf 15}M} = \partial_M +2ig_5 A_M+i\frac{2}{3}g_5' B_M . 
\end{equation}
The subscripts $t$ and $b$ in the $\vec T$ and $\vec B$ mean 
 that they have the same electric charge of $t$ and $b$ quarks, respectively. 
$T_{3,15}$ and $B_{3,15}$ mean $SU(2)$ quark doublets involved 
 in the $\bf 3$ and $\overline{\bf 15}$ representations. 
The W boson mass is $M_W=gv/2$.
The matrices $M_t$ and $M_b$ are defined by
\begin{equation}
M_t=
\begin{pmatrix}
0&0&2&0&0&0\\
0&0&0&0&0&0\\
2&0&0&\sqrt6&0&0\\
0&0&\sqrt6&0&\sqrt6&0\\
0&0&0&\sqrt6&0&2\\
0&0&0&0&2&0
\end{pmatrix}
, \quad
M_b=
\begin{pmatrix}
0&1&0&0&0&0\\
1&0&0&0&0&0\\
0&0&0&\sqrt3&0&0\\
0&0&\sqrt3&0&2&0\\
0&0&0&2 &0 &\sqrt3\\
0&0&0&0&\sqrt3&0
\end{pmatrix}.
\end{equation}
The boundary conditions are
\begin{equation}
\label{eq:BCmessenger}
\begin{cases}
\vec T(-y)=-\gamma_5P\vec T(+y)\\
\vec T(y)=-\vec T(y+2\pi R)
\end{cases}
,
\begin{cases}
\vec B(-y)=-\gamma_5P\vec B(+y)\\
\vec B(y)=-\vec B(y+2\pi R)
\end{cases}
\end{equation}
where $P=\text{diag}(+,-,-,+,-,+)$.

We first focus on the top quark KK mass spectrum. 
The equations of motion (EOM) of $t$ quark are
\begin{align}
\label{EOM_top}
\dis
&0=i\partial_\mu\Gamma^\mu \vec T_L
+i\partial_5\Gamma^5 \vec T_R
+i M_W M_t \Gamma^5\vec T_R
+\frac{\epsilon_{tR}}{\sqrt{\pi R}}
(t_R,0,0,0,0,0)^\T
\delta(y-\pi R),
\\
\dis
&0=i\partial_\mu\Gamma^\mu \vec T_R
+i\partial_5\Gamma^5 \vec T_L
+i M_W M_t\Gamma^5 \vec T_L
+\frac{\epsilon_L}{\sqrt{\pi R}}
(0,\sin\theta t_L,\cos\theta t_L,0,0,0)^\T
\delta(y-\pi R),
\\
\dis
&0=\left[i\partial_\mu\gamma^\mu t_L
+\frac{\epsilon_L}{\sqrt{\pi R}}(\cos\theta T_{15R}+\sin\theta T_{3R})\right]\delta(y-\pi R),
\\
\dis
&0=\left[i\partial_\mu\gamma^\mu t_R
+\frac{\epsilon_{tR}}{\sqrt{\pi R}}T_L\right]\delta(y-\pi R).
\end{align}
The field redefinition
\begin{equation}
\vec {\tilde T}=\exp[iM_W M_t y]\vec T
\end{equation}
simplifies the bulk equations as 
\begin{align}
&0=i\partial_\mu\gamma^\mu \vec {\tilde T}_L
-\partial_y \vec {\tilde T}_R,
\\
&0=i\partial_\mu\gamma^\mu \vec {\tilde T}_R
+\partial_y \vec {\tilde T}_L.
\end{align}
Then we obtain the following mode functions respecting the  $Z_2$ parities 
at $y=0$.
\begin{align}
\vec {\tilde T}_L
\propto
\frac{1}{\sqrt{\pi R}}
\begin{bmatrix}
\sin (m_ny) T_L^n\\
\cos (m_ny) T_{3L}^n\\
\cos (m_ny) T_{15L}^n\\
\sin (m_ny) \Sigma_{tL}^n\\
\cos (m_ny) \Lambda_{tL}^n\\
\sin (m_ny) \Delta_{tL}^n\\
\end{bmatrix}
, \quad 
\vec {\tilde T}_R
\propto
\frac{1}{\sqrt{\pi R}}
\begin{bmatrix}
\cos (m_ny) T_R^n\\
-\sin (m_ny) T_{3R}^n\\
-\sin (m_ny) T_{15R}^n\\
\cos (m_ny) \Sigma_{tR}^n\\
-\sin (m_ny) \Lambda_{tR}^n\\
\cos (m_ny) \Delta_{tR}^n\\
\end{bmatrix}.
\end{align}
$m_n$ stands for mass eigenvalues:
$i\partial_\mu\gamma^\mu\vec {T}=m_n\vec {T}$.
In order to obtain the mass spectrum, 
 we have to impose the boundary conditions. 
One is an anti-periodic boundary condition with respect to $S^1$
and the other is 
 the boundary condition at the $y=\pi R$
 that is precisely discussed in \cite{Csaki:2003sh}. 
The latter can be obtained by integrating out the EOM around $y=\pi R$ 
:
\begin{equation}
0=
\lim_{\varepsilon\to 0}
\int_{\pi R-\varepsilon}^{\pi R}\rmd y [\text{EOM}].
\end{equation}
For example, the boundary condition
 from the first line of eq.(\ref{EOM_top}) gives 
\begin{align}
0=&
\lim_{\varepsilon\to 0}
\int_{\pi R-\varepsilon}^{\pi R}\rmd y 
\left[
i\partial_\mu\Gamma^\mu \vec T_L
+i\partial_5\Gamma^5 \vec T_R
+i M_W M_t \Gamma^5\vec T_R
+\frac{\epsilon_{tR}}{\sqrt{\pi R}}
(t_R,0,0,0,0,0)^\T
\delta(y-\pi R)
\right]
\nonumber \\
=&
\lim_{\varepsilon\to 0}
\left[
i\Gamma^5 
(T_R,T_{3R},T_{15R},\Sigma_{tR},\Lambda_{tR},\Delta_{tR})^\T
\right]_{\pi R-\varepsilon}^{\pi R}
+\frac{\epsilon_{tR}}{\sqrt{\pi R}}
(t_R,0,0,0,0,0)^\T
\end{align}
where we use the fact that the bulk fields are continuous.
From the $Z_2$ parities at the $y=\pi R$
which are derived from eq.(\ref{eq:BCmessenger}), 
 $Z_2$ odd fields vanish at the fixed point: $0=T_R(\pi R)=\Sigma_{tR}(\pi R)=\Delta_{tR}(\pi R)$.
Simplifying the notation as $\pi R^- = \pi R -\varepsilon$,
 it becomes
\begin{align}
\label{derivedBC_tR}
0=&
i\Gamma^5 
\lim_{\varepsilon\to 0}
\begin{pmatrix}
-T_R(\pi R^-)\\
T_{3R}(\pi R)-T_{3R}(\pi R^-)\\
T_{15R}(\pi R)-T_{15R}(\pi R^-)\\
-\Sigma_{tR}(\pi R^-)\\
\Lambda_{tR}(\pi R)-\Lambda_{tR}(\pi R^-)\\
-\Delta_{tR}(\pi R^-)
\end{pmatrix}
+\frac{\epsilon_{tR}}{\sqrt{\pi R}}
\begin{pmatrix}
t_R\\0\\0\\0\\0\\0
\end{pmatrix}. 
\end{align}
The first component indicates that 
 the boundary conditions are modified 
 by the boundary term.
The others are ordinary boundary conditions:
 continuity conditions and $Z_2$ conditions.
Combining the first relation in (\ref{derivedBC_tR}) and the EOM for the boundary fermion $t_R$ (\ref{EOM_top}),
 we have
\begin{equation}
0=
\frac{\epsilon_{tR}}{\sqrt{\pi R}}i\partial_\mu\gamma^\mu t_R
+\frac{\epsilon_{tR}^2}{{\pi R}}T_L
=
i\partial_\mu\gamma^\mu T_R
+\frac{\epsilon_{tR}^2}{{\pi R}}T_L. 
\end{equation}
To summarize, the modified boundary conditions are
\begin{align}
\dis
&0=i\partial_\mu\gamma^\mu T_R(\pi R^-)
-\frac{\epsilon_{tR}^2}{\pi R} T_L(\pi R^-),
\\
\dis
&0=i\partial_\mu\gamma^\mu T_{3L}(\pi R^-)
+\frac{\epsilon_L^2}{\pi R}\sin\theta
[\cos\theta T_{15R}(\pi R^-)+\sin\theta T_{3R}(\pi R^-)],
\\
\dis
&0=i\partial_\mu\gamma^\mu T_{15L}(\pi R^-)
+\frac{\epsilon_L^2}{\pi R}\cos\theta
[\cos\theta T_{15R}(\pi R^-)+\sin\theta T_{3R}(\pi R^-)],
\\
\dis
&0=\Sigma_{tR}(\pi R^-)=\Lambda_{tL}(\pi R^-)
=\Delta_{tR}(\pi R^-).
\end{align}
Taking into account these boundary conditions,
 we find a very complicated relation determining the KK mass spectrum of the top quark.
\begin{align}
\label{massTOP}
0=&
 2\hat m_n^2\cos^2\hat m_n\left(\cos^2\hat m_n-\sin^2(2\hat M_W)\right)
 \left(\cos^2\hat m_n-\sin^2(4\hat M_W)\right)
\nonumber \\
&-\epsilon_L^2\hat m_n\cos \hat m_n\sin \hat m_n
\nonumber \\
&
 \times  \Big[
 \sin^2\theta\left\{\sin^2(4 \hat M_W) \cos^2\hat m_n 
 +\sin^2 (2 \hat M_W) \cos^2 \hat m_n
 -2\sin^2 (2\hat M_W) \sin^2 (4 \hat M_W)
 \right\}
 \nonumber \\
 &
 -2\cos^4\hat m_n+\sin^2(4\hat M_W) \cos^2\hat m_n
 +\sin^2(2\hat M_W) \cos^2 \hat m_n
 \Big]
\nonumber \\
&
 +\frac{\epsilon_{tR}^2}{4}\hat m_n\sin \hat m_n\cos \hat m_n
\nonumber  \\
&
 \times \Big[
 8\cos^4\hat m_n
 -7\sin^2(4\hat M_W)\cos^2\hat m_n
 -4\sin^2(2\hat M_W)\cos^2\hat m_n
 +3\sin^2(2\hat M_W)\sin^2(4\hat M_W)
 \Big]
\nonumber  \\
&-\frac{\epsilon_L^2\epsilon_{tR}^2}{8}
\Big[
 \cos^2\theta
 \Big\{
  8(\sin^2(4\hat M_W)+\sin^2(2\hat M_W))\cos^4\hat m_n
 \nonumber \\
  &
  +\cos^2\hat m_n
  \left(
    (-11\sin^2(2\hat M_W)-7)\sin^2(4\hat M_W)
    +\sin(4\hat M_W)\sin(8\hat M_W)
    -4\sin^2(2\hat M_W)\right)
  \nonumber \\
  &+6\sin^2(2\hat M_W)\sin^2(4\hat M_W)
 \Big\}
 +16\cos^6\hat m_n
 -2(7\sin^2(4\hat M_W)+4\sin^2(2\hat M_W)+8)\cos^4\hat m_n
 \nonumber \\
 &~~
 +2\cos^2\hat m_n
 \left\{
 (3\sin^2(2\hat M_W)+7)\sin^2(4\hat M_W)+4\sin^2(2\hat M_W)
 \right\}
 -6\sin^2(2\hat M_W)\sin^2(4\hat M_W)
\Big] 
\end{align}
where $\hat m_n=\pi Rm_n$ and $\hat M_W=\pi R M_W$ are dimensionless parameters 
 normalized by $\pi R$.
The lightest mass eigenvalue can be found 
 by taking the limit $\hat M_W\to 0$:
\begin{equation}
\hat m_t^2
=
 \frac{4\epsilon_L^2\epsilon_{tR}^2}{(1+\epsilon_L^2)(1+\epsilon_{tR}^2)}(1-\sin^2\theta)\hat M_W^2 
+\mathcal O (\hat M_W^4) \,.
\end{equation}
For the small $\theta$ and large $\epsilon_{tR}$ and $\epsilon_L$, the $m_t$ almost equal to $2 M_W$.
This result  allows us to interpret
it as top quark.

 The bottom quark mass is obtained by the same procedure as top quark mass 
 except for the yukawa coupling.
The EOM for the bottom quark sector become
\begin{align}
\dis
&0=i\partial_\mu\Gamma^\mu \vec B_L
+i\partial_5\Gamma^5 \vec B_R
+i M_W M_b \Gamma^5\vec B_R
+\frac{\epsilon_{bR}}{\sqrt{\pi R}}
(
b_R,0,0,0,0,0
)^\T
\delta(y-\pi R),
\\
\dis
&0=i\partial_\mu\Gamma^\mu \vec B_R
+i\partial_5\Gamma^5 \vec B_L
+iM_W M_b\Gamma^5 \vec B_L
+\frac{\epsilon_L}{\sqrt{\pi R}}
(
0,\sin\theta b_L,\cos\theta b_L,0,0,0
)^\T
\delta(y-\pi R),
\\
\dis
&0=\left[i\partial_\mu\gamma^\mu b_L
+\frac{\epsilon_L}{\sqrt{\pi R}}(\cos\theta B_{15R}+\sin\theta B_{3R})\right]\delta(y-\pi R),
\\
\dis
&0=\left[i\partial_\mu\gamma^\mu b_R
+\frac{\epsilon_{bR}}{\sqrt{\pi R}}B_L\right]\delta(y-\pi R).
\end{align}
The mode functions are given as
\begin{align}
&\vec {\tilde B}_L
\propto
\frac{1}{\sqrt{\pi R}}
\begin{bmatrix}
\sin (m_ny) B_L^n\\
\cos (m_ny) B_{3L}^n\\
\cos (m_ny) B_{15L}^n\\
\sin (m_ny) \Sigma_{bL}^n\\
\cos (m_ny) \Lambda_{bL}^n\\
\sin (m_ny) \Delta_{bL}^n\\
\end{bmatrix}
,\quad
\vec {\tilde B}_R
\propto
\frac{1}{\sqrt{\pi R}}
\begin{bmatrix}
\cos (m_ny) B_R^n\\
-\sin (m_ny) B_{3R}^n\\
-\sin (m_ny) B_{15R}^n\\
\cos (m_ny) \Sigma_{bR}^n\\
-\sin (m_ny) \Lambda_{bR}^n\\
\cos (m_ny) \Delta_{bR}^n\\
\end{bmatrix}
, \\
&\vec B =\exp[iM_WM_b y] \vec {\tilde B}. 
\end{align}
The modified boundary conditions are 
\begin{align}
\dis
&0=i\partial_\mu\gamma^\mu B_R(\pi R^-)
-\frac{\epsilon_{tR}^2}{\pi R} B_L(\pi R^-),
\\
\dis
&0=i\partial_\mu\gamma^\mu B_{3L}(\pi R^-)
+\frac{\epsilon_L^2}{\pi R}\sin\theta
[\cos\theta B_{15R}(\pi R^-)+\sin\theta B_{3R}(\pi R^-)],
\\
\dis
&0=i\partial_\mu\gamma^\mu B_{15L}(\pi R^-)
+\frac{\epsilon_L^2}{\pi R}\cos\theta
[\cos\theta B_{15R}(\pi R^-)+\sin\theta B_{3R}(\pi R^-)],
\\
\dis
&0=\Sigma_{bR}(\pi R^-)
=\Lambda_{bL}(\pi R^-)
=\Delta_{bR}(\pi R^-).
\end{align}
Repeating the same analysis, 
 we find a corresponding relation determining the KK mass spectrum of the bottom quark. 
\begin{align}
\label{massBOTTOM}
0=&
-2\hat m_n^2(\sin^2\hat m_n-\cos^2\hat M_W)^2(\sin^2\hat m_n-\cos^2(3\hat M_W)) 
\nonumber \\
&
-\frac{\epsilon_L^2}{2} \hat m_n\sin\hat m_n\cos\hat m_n
(\sin^2\hat m_n-\cos^2\hat M_W)
\nonumber \\
&
\times \left[
\sin^2\theta(\cos^2(3\hat M_W)-\cos^2\hat M_W)
-4\sin^2(\hat m_n)+3\cos^2(3\hat M_W)+\cos^2(\hat M_W)
\right]
\nonumber \\
&
+2\epsilon_{bR}^2\hat m_n\sin\hat m_n\cos\hat m_n
 (\sin^2\hat m_n-\cos^2\hat M_W)
 (\sin^2\hat m_n-\cos^2(3\hat M_W))
 \nonumber \\
&
+\frac{\epsilon_L^2\epsilon_{bR}^2}{2}
\Big[
\Big\{(\cos^2\hat M_W-\cos^2(3\hat M_W))\sin^4\hat m_n 
-4\sin^2\hat M_W\cos^2\hat M_W\cos^2(3\hat M_W)
\nonumber \\
&
+ (\cos^2(3\hat M_W)-4\cos^4\hat M_W+3\cos^2\hat M_W)\sin^2\hat m_n
\Big\}\sin^2\theta
+4\sin^6\hat m_n 
\nonumber \\
&
- \left(3\cos^2(3\hat M_W)+\cos^2\hat M_W+4\right)\sin^4\hat m_n+(3\cos^2(3\hat M_W)+\cos^2(\hat M_W))\sin^2\hat m_n
\Big]
\end{align}
The lightest mass is obtained as
\begin{equation}
\hat m^2_b=
\frac{\epsilon_L^2\epsilon_{bR}^2}{(1+\epsilon_{L}^2)(1+\epsilon_{tR}^2)}\sin^2\theta \hat M_W^2 +\mathcal O(\hat M_W^4)
\label{bottom}
\end{equation}
where $\hat m_b=\pi Rm_b$. 
The bottom mass can be achieved by tuning parameters $\theta,\epsilon_{bR}$.
To reproduce the observed masses 
 $m_t=170{\rm GeV},m_b=4.2{\rm GeV}$ in (\ref{bottom}),
 $\epsilon_{L}, \epsilon_{tR} \gg 1$ and $\theta \ll 1$ are required.
For example, if we choose 
$\epsilon_L=\epsilon_{tR}=10,\theta=0.1,\epsilon_{bR}=0.6$, 
 then we obtain $m_t\sim 1.97M_W\sim 158{\rm GeV}, 
m_b\sim 4.09{\rm GeV}$. 

We note that the exotic fermions with the different quantum numbers 
 from those of SM particle are included in the $\overline{\bf 15}$ representation.
Their spectrum are given by the solutions of the following equations
\begin{align}
\label{massexotic}
&0=\cos \hat m_n \cos(\hat m_n-2 \hat M_W) \cos( \hat m_n + 2 \hat M_W),
\nonumber \\
&0=\cos(\hat m_n- \hat M_W) \cos(\hat m_n+ \hat M_W),
\\
&0=\cos \hat m_n. 
\nonumber 
\end{align}
The lightest mode of the exotic fermions obtain a mass around $\sim 1/(2R)$
 as we mentioned before.


\subsection{The first two generations of quarks and three generations of leptons}
In this subsection, 
 we derive the mass spectrum of the $\bf 3$ and $ \overline{\bf 3}$ representation.
Since the procedure is almost similar, 
 we only point out the differences.
The $Q_\H$ and $L_\H$ are given by 
\begin{equation}
(Q_\SM,Q_\H)^\T = U_Q (Q_3,Q_{\bar 3})^\T
, \quad
(L_\SM,L_\H)^\T = U_l (L_3,L_{\bar 3})^\T.
\end{equation}
The $Q_\H$ and $L_\H$ has the brane mass term and 
 become massive.
On the other hand,
 the $Q_\SM$ and $L_\SM$ are left massless 
 which correspond to the SM doublets.

The EOM of the down-type quark is derived as
\begin{align}
\label{EOM}
  \left[i\partial_M\Gamma^M +iM_W\sigma_1 \gamma^5 
  -M\epsilon(y)\right]\vec d
  &= -\frac{\epsilon}{\sqrt{\pi R}} U_Q^\dag (0,0,d_\B)^\T\delta(y),
  \\
  i\partial_\mu \gamma^\mu d_\B\delta(y)
  &=-\frac{\epsilon}{\sqrt{\pi R}}\delta(y) (U_Q \vec d)_3 
  =-\frac{\epsilon}{\sqrt{\pi R}} \delta(y) d_{\rm H}
\end{align}
where $\vec d =(d,d_3,d_{\bar 3})^{\rm T}$.
The mixing matrix $U_Q$ is defined by
\begin{equation}
  U_Q=\begin{pmatrix}
   1&0&0\\0&\cos\theta_q&-\sin\theta_q\\0&\sin\theta_q&\cos\theta_q
  \end{pmatrix}.
\end{equation}
The KK expansions of bulk fields are given by
solving the bulk equation and respecting the $Z_2$ parities:
\begin{align}
&\vec {\tilde d}_L\propto \frac{1}{\sqrt{\pi R}}
\begin{bmatrix}
\sin\left(\sqrt{m_n^2-M_q^2}y\right)d^{n}_L
\\
\left\{
\sqrt{\hat m_n^2-\hat M_q^2}\cos(\sqrt{m_n^2-M_q^2}y)
+\hat M_q\sin(\sqrt{m_n^2-M_q^2}|y|)
\right\}d^{n}_{\SM L}
\\
\left\{
\sqrt{\hat m_n^2-\hat M_q^2}\cos(\sqrt{m_n^2-M_q^2}y)
+(\hat M_q+\epsilon_q^2/2)\sin(\sqrt{m_n^2-M_q^2}|y|)
\right\}d^{n}_{\H L}
\end{bmatrix},
\\
&\vec {\tilde d}_R\propto\frac{1}{\sqrt{\pi R}}
\begin{bmatrix}
\left\{
\sqrt{\hat m_n^2-\hat M_q^2}\cos(\sqrt{m_n^2-M_q^2}y)
-\hat M_q\sin(\sqrt{m_n^2-M_q^2}|y|)
\right\}
d^{n}_R 
\\
\sin (\sqrt{m_n^2-M_q^2}y)d^{n}_{\SM R}
\\
\sin (\sqrt{m_n^2-M_q^2}y)d^{n}_{\H R}
\end{bmatrix}.
\end{align}
They are determined to satisfy the boundary condition at the origin.
  \begin{equation}
   \lim_{\varepsilon\to 0}
   \int_{-\varepsilon}^{\varepsilon} \rmd y
   \left[\text{EOM}\right]
   =0.  
  \end{equation}
We impose the periodic boundary condition on the bulk fermion
\begin{equation}
 \label{dmasscond1}
 0=[\vec d (y)]_{y=-\pi R}^{y=\pi R}
\end{equation}
and the boundary conditions at $y=\pm \pi R$
,
we have
\begin{equation}
 \label{dmasscond2}
 0=[\text{EOM}]_{y=-\pi R}^{y=\pi R}. 
\end{equation}
It gives the conditions on the first derivative of mode function.

The KK mass spectrum is obtained from the eqs.(\ref{dmasscond1}) and (\ref{dmasscond2}).
Substituting the mode expansion into these conditions,
 we find the KK mass spectrum for the down-type quark 
 by solving a equation: 
\begin{align}
0&=
\sqrt{\hat m_n^2-\hat M_q^2}\epsilon^2_q
\left[(\hat M_q^2-\hat m_n^2)\sin^2\hat M_W \cos^2\sqrt{\hat m_n^2-\hat M_q^2}\sin^2\theta_q
+\hat m_n^2
\sin^2\sqrt{\hat m_n^2-\hat M_q^2}\right]
\nonumber \\
&\times \cos\sqrt{\hat m_n^2-\hat M_q^2}
 +\epsilon^2_q\sin^2\theta_q\hat M_q(\hat m_n^2-\hat M_q^2)\sin^2\hat M_W\sin\sqrt{\hat m_n^2-\hat M_q^2}
\nonumber \\&
 +\hat m_n^2\left[(\hat M_q\epsilon^2_q-2\hat m_n^2)\cos^2\sqrt{\hat m_n^2-\hat M_q^2}
 +2\hat m_n^2\cos^2\hat M_W+\hat M_q(2\hat M_q\sin^2\hat M_W-\epsilon^2_q)\right]
\nonumber \\
&\times \sin\sqrt{\hat m_n^2-\hat M_q^2}.
\end{align}
As for the up-type quark,
 the KK mass spectrum can be found by solving a equation: 
\begin{align}
0=&
\sqrt{\hat m_n^2-\hat M_q^2}\epsilon^2_q\left[(\hat M_q^2-\hat m_n^2)\sin^2\hat M_W\cos^2\theta_q+\hat m_n^2\sin^2\sqrt{\hat m_n^2-\hat M_q^2}\right]\cos\sqrt{\hat m_n^2-\hat M_q^2}
\nonumber \\&
 +\hat M_q\epsilon_q^2(\hat m_n^2-\hat M_q^2)\sin^2\hat M_W\sin\sqrt{\hat m_n^2-\hat M_q^2}\cos^2\theta_q
\nonumber \\&
 +\hat m_n^2(\hat m_n^2-\hat M_q\epsilon_q^2)\sin^3\sqrt{\hat m_n^2-\hat M_q^2}
 +2\hat m_n^2(\hat M_q^2-\hat m_n^2)\sin^2\hat M_W\sin\sqrt{\hat m_n^2-\hat M_q^2}.
\end{align}
The lightest masses can be obtained as
\begin{equation}
\hat m_u^2
=
 \frac{\hat M_q^2}{\sinh^2\hat M_q}\sin^2\theta_q M_W^2
 +\mathcal O (\hat M_W^4)
 , \quad
\hat m_d^2
=
 \frac{\hat M_q^2}{\sinh^2\hat M_q}\cos^2\theta_q M_W^2
 +\mathcal O (\hat M_W^4). 
\end{equation}
This result is easily understood from the fact that 
 the angle $\theta_q$ represents how the singlets in the each representations 
 couple to the SM doublet $Q_\SM$.
Namely, the SM doublet is purely $Q_{\bf 3}(Q_{\bf{\bar 3}})$ in the case $\theta =0(\pi/2)$, 
 so the singlet $u(d)$ in the $\bar{\bf 3}({\bf 3})$ cannot connect to the SM doublet.   
The lepton sector is completely the same in this scenario. 
We can read the lepton masses from the above result 
 by replacing $M_q\to M_l,\epsilon_q\to\epsilon_l$ and $\theta_q\to\theta_l$.


\subsection{Mirror fermion}
As will be seen in the next section,
 the dominant contributions from fermions with the anti-periodic boundary condition 
 to the Higgs potential at 1-loop behave as bosonic fields, 
 which implies that the contributions from the extra bulk fermions with periodic boundary condition 
 are indispensable for realizing the realistic electroweak symmetry breaking.
To accomplish it,
 we introduce two massive fermions
$\Psi_\M$ and $X_\M$ 
 with the relative opposite $Z_2$ parities, 
 which we call mirror fermions.

To investigate the spectrum of the mirror fermion,
we begin with the triplet mirror fermion as the simplest example:
\begin{equation}
\Psi_\M=\begin{pmatrix}
\Psi_1\\ \Psi_2 \\ \Psi_3
\end{pmatrix}
, \quad
X_\M=\begin{pmatrix}
X_1\\ X_2 \\ X_3
\end{pmatrix}
\end{equation}
Since the first components does not couple with the Higgs boson, 
 we concentrate on the lower two components.

Hereafter, the vector notation $\vec \Psi_\M=(\Psi_2,\Psi_3)^\T$ and 
 $\vec X_\M=(X_2,X_3)^\T$ are employed, and then the EOM becomes
\begin{align}
&0
=i\partial_\mu\gamma^\mu \vec \Psi_\M +i\partial_y\Gamma^5 \vec \Psi_\M-M\vec X_\M +\Gamma^5M_W\sigma_1 \vec \Psi_\M
\\
&0
=i\partial_\mu\gamma^\mu \vec X_\M +i\partial_y\Gamma^5 \vec X_\M-M\vec \Psi_\M +\Gamma^5M_W\sigma_1 \vec X_\M 
\end{align}
where $\sigma_1$ is a Pauli matrix.
Eliminating $M_W$ by the 
field redefinition 
as
\begin{equation}
\vec {\tilde \Psi}_\M= \e^{-iM_W\sigma_1 y}\vec \Psi_\M, \quad 
\vec {\tilde X}_\M= \e^{-iM_W\sigma_1 y}\vec X_\M,
\end{equation}
we have
\begin{align}
&0
=i\partial_\mu\gamma^\mu \vec {\tilde \Psi}_\M 
+i\partial_y\Gamma^5 \vec {\tilde \Psi}_\M
-M\vec {\tilde X}_\M, 
\\
&0
=i\partial_\mu\gamma^\mu \vec {\tilde X}_\M 
+i\partial_y\Gamma^5 \vec {\tilde X}_\M
-M\vec {\tilde \Psi}_\M.  
\end{align}
In this base,
 the bulk equations are easily solved as
\begin{align}
\vec {\tilde \Psi}_\M
=&
\frac{1}{\sqrt{\pi R}}
\begin{bmatrix}
 \psi_{2L}^n\cos\sqrt{m_n^2-M^2}y+\psi_{2R}^n\sin\sqrt{m_n^2-M^2}y
 \\
 \psi_{3R}^n\cos\sqrt{m_n^2-M^2}y+\psi_{3L}^n\sin\sqrt{m_n^2-M^2}y
 \end{bmatrix}
 ,
 \\
\vec {\tilde X}_\M
=&
\frac{1}{\sqrt{\pi R}}
\begin{bmatrix}
 \chi_{2L}^n\sin\sqrt{m_n^2-M^2}y+\chi_{2R}^n\cos\sqrt{m_n^2-M^2}y
 \\
 \chi_{3R}^n\sin\sqrt{m_n^2-M^2}y+\chi_{3L}^n\cos\sqrt{m_n^2-M^2}y
\end{bmatrix}.
\end{align}
From the periodic boundary conditions 
 and the EOM at the fixed points
\begin{align}
0=[\vec \Psi_\M]_{-\pi R}^{+\pi R}, \quad 
0=
[\text{EOM}]_{-\pi R}^{+\pi R}
\propto[\partial_y \vec \Psi_\M]_{-\pi R}^{+\pi R},
\end{align}
the KK mass spectrums are obtained from
\begin{equation}
0
=\sin\left(\sqrt{\hat m_n^2-\hat M^2}-\hat M_W\right)
\sin\left(\sqrt{\hat m_n^2-\hat M^2}+\hat M_W\right). 
\end{equation}
%
Noting that the bulk mass for the extra fermions are constrained 
 from the search for the fourth generation fermions, 
 the mass of the lightest mode in the extra bulk fermion
 should be larger than the $\mathcal O(700 \rm GeV)$ or so \cite{Patrignani:2016xqp}, 
 which implies that the bulk mass of the extra bulk fermion
 must satisfy the lower bound
\begin{equation}
M>\sqrt{(700{\rm GeV})^2-M_W^2}. 
\end{equation}



\section{Higgs potential analysis}

Now, we are ready to discuss the Higgs potential generated by the quantum corrections.
Since some of the mass spectrum cannot be solved explicitly,
 we employ the $\zeta$ function regularization method.
A particle with the mass $m_n$ contributes to the 1-loop effective potential as follows. 
\begin{equation}
 \label{EP1}
 V_\text{5D}
 =
\frac{1}{2\pi R} \int\frac{\rmd ^4p_{\rm E}}{(2\pi)^4}
 \frac{(-1)^FN_\text{DOF}}{2}\sum_{n=-\infty}^{\infty}
 \ln\left(p_{\rm E}^2+m_n^2\right)
\end{equation}
where the $N_\text{DOF}$  stands for the degree of freedom and 
 $F=1(0)$ for the fermion (boson).
The above infinite summation can be rewritten by the following integral form as
\begin{align}
 \label{EP_integral_form}
  V_\text{5D} 
 =&
 -\frac{1}{2\pi R} 
 \frac{(-1)^FN_\text{DOF}}{32\pi^2}
 \frac{1}{R^4}
 \int_0^\infty \rmd u~u^{4}\frac{\rmd}{\rmd u}\ln\left[N(iu)\right]
\end{align}
The mass spectrum $m_n$ is determined 
by zeros of the function $N(iu)$, 
\begin{equation}
N(m_n)=0.
\end{equation}
The function $N(iu)$ is defined as such that $\hat m_n$ and $\hat M_W$ are replaced by 
 $i \pi u$ and $\pi \alpha$ in the relation determining the KK mass spectrum, respectively. 
As an illustration, the functions $N_W(iu)$ and $N_Z(iu)$ for $W$ and $Z$ gauge bosons 
 are explicitly shown. 
\begin{align}
N_W(iu)=&\cosh^2(\pi u)-\cos^2(\pi\alpha),
\\
N_Z(iu)
=&
-\tanh^2(\pi u)-
\frac{\sin^2(\pi\alpha)[4\cos^2\theta_W-\sin^2(\pi\alpha)]}
{(2\cos^2\theta_W-\sin^2(\pi\alpha))^2}
\end{align}
where $\theta_W$ is the weak mixing angle.
One can verify that these functions are obtained by the above replacements 
 in the relations determining the KK spectrum of $W$ and $Z$ bosons \cite{Adachi:2016zdi}, 
\begin{align}
0=& \cos^2 (\hat m_n) - \cos^2(\hat M_W),
\\
0
=&
\tan^2(\hat m_n) -
\frac{\sin^2(\hat M_W)[4\cos^2\theta_W-\sin^2(\hat M_W)]}
{(2\cos^2\theta_W-\sin^2(\hat M_W))^2}. 
\end{align}
The four dimensional effective potential 
is given by integrating out the extra dimension:
\begin{equation}
V= \int_0^{2\pi R}\rmd y~ V_\text{5D}
=- 
 \frac{(-1)^FN_\text{DOF}}{32\pi^2}
 \frac{1}{R^4}
 \int_0^\infty \rmd u~u^{4}\frac{\rmd}{\rmd u}\ln\left[N(iu)\right]. 
\end{equation}
Finally, the 1-loop Higgs effective potential of our model is given by 
\begin{align}
V_\text{R}
=&
 -
 \frac{1}{32\pi^2}
 \frac{1}{R^4}
 \int_0^\infty \rmd u\,u^{4}
 \frac{\rmd}{\rmd u}
 \Big[
 3\ln N_Z(iu)+3\ln N_W(iu)
 \nonumber \\
 &
 -3\cdot4\ln N_\text{BOT}(iu)
 -3\cdot4\ln N_\text{TOP}(iu)
 -3\cdot4\ln N_\text{exotic}(iu)
 -3\cdot4\ln N_\text{M}(iu)
 \Big]
 -(\alpha\to 0)
  \label{effpot}
\end{align}
where $N_\text{BOT}(iu)$, $N_\text{TOP}(iu)$, $N_\text{exotic}(iu)$ and $N_\text{M}(iu)$ 
 are the functions for the bottom quark, top quark, the exotic fermions and the mirror fermions. 
Their explicit forms are omitted since they are very lengthy and complicated. 
These functions can be similarly obtained like $N_{W/Z}(iu)$ as explained above. 
Note that the divergent $\alpha$ independent terms in the effective potential, 
 which are vacuum energy, are subtracted. 
  
\begin{figure}[h]
\begin{center}
\includegraphics[scale=0.4]{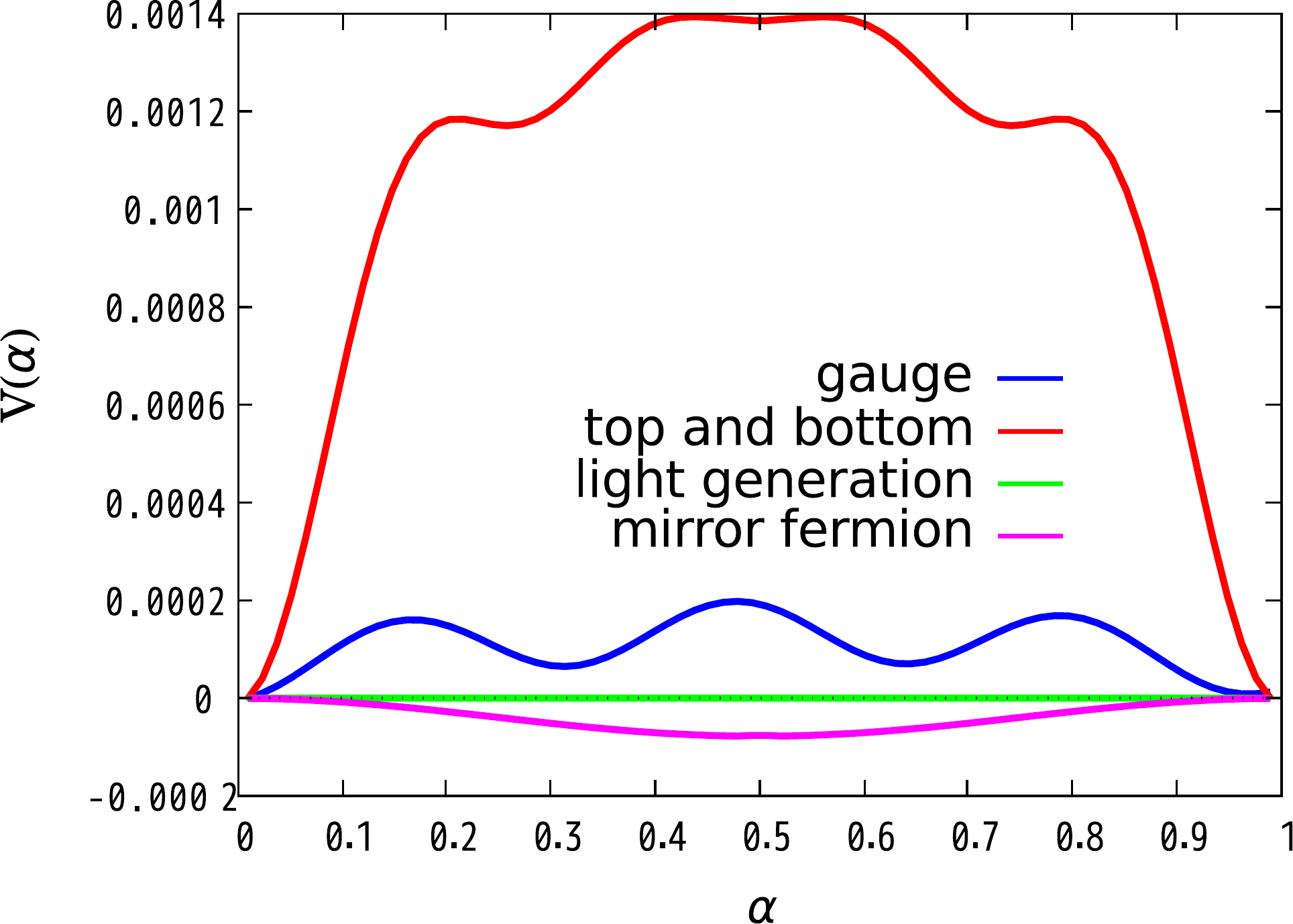}
\end{center}
\caption{Contributions from the gauge bosons, the SM quarks, exotic and mirror fermion.
The mirror fermion in this figure is {\bf 3} representation.
}
\label{fig:EPSMandmirror}
\end{figure}

Let us discuss the behavior of the effective potential in detail. 
First, the effective potential from the SM fields and messenger fermions 
 is shown in Figure \ref{fig:EPSMandmirror}.
We immediately see that the third generation of quarks give 
 dominant contributions to the effective potential 
 since they have no bulk mass term. 
Note that the contributions from the third generation behave as bosonic field 
 similar to the gauge fields due to the anti-periodicity. 
In particular, the potential curvature at the origin is positive.  
As for other SM fermions, they have bulk mass terms and the yukawa couplings are highly suppressed
 by the factor $\e^{-\pi RM_q}$ or  $\e^{-\pi RM_l}$ and
 therefore their contributions to the effective potential are negligible and 
 will not be included in the potential analysis later.

\begin{figure}[h]
\begin{center}
\includegraphics[scale=0.4]{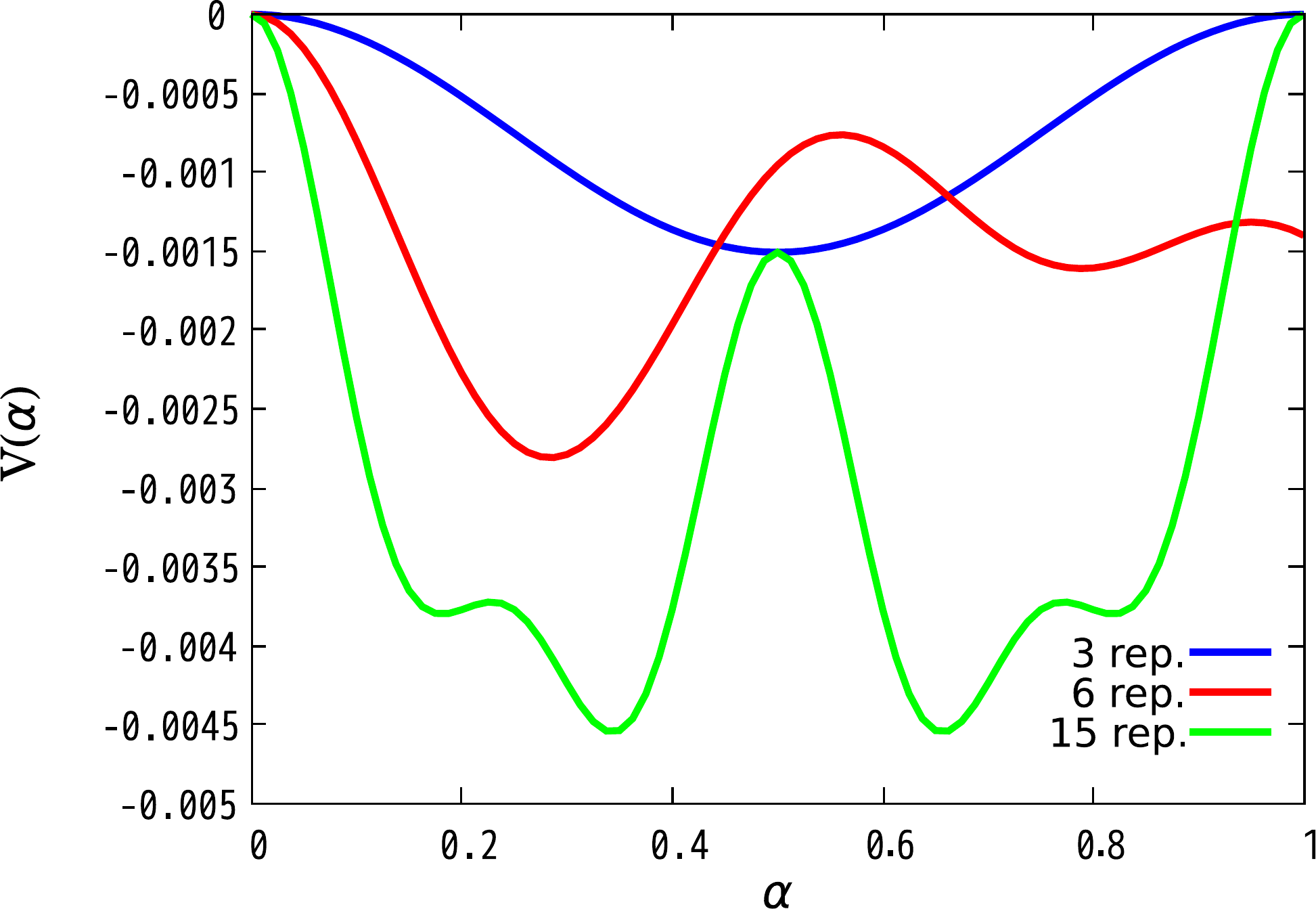}
\end{center}
\caption{Effective potential of mirror fermions.}
\label{fig:mirror}
\end{figure}

\begin{figure}[h]
\begin{center}
\includegraphics[scale=0.35]{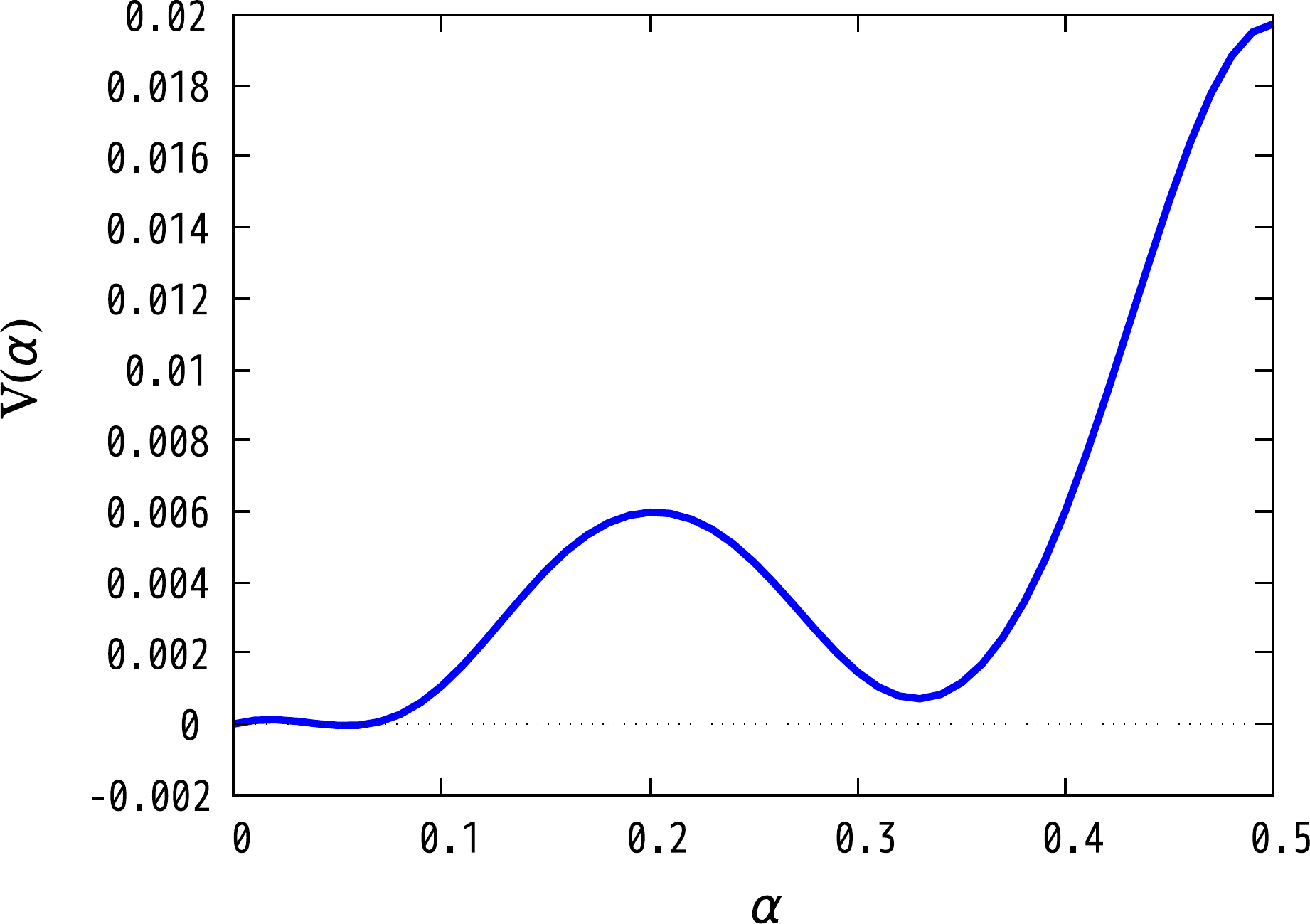}
\includegraphics[scale=0.35]{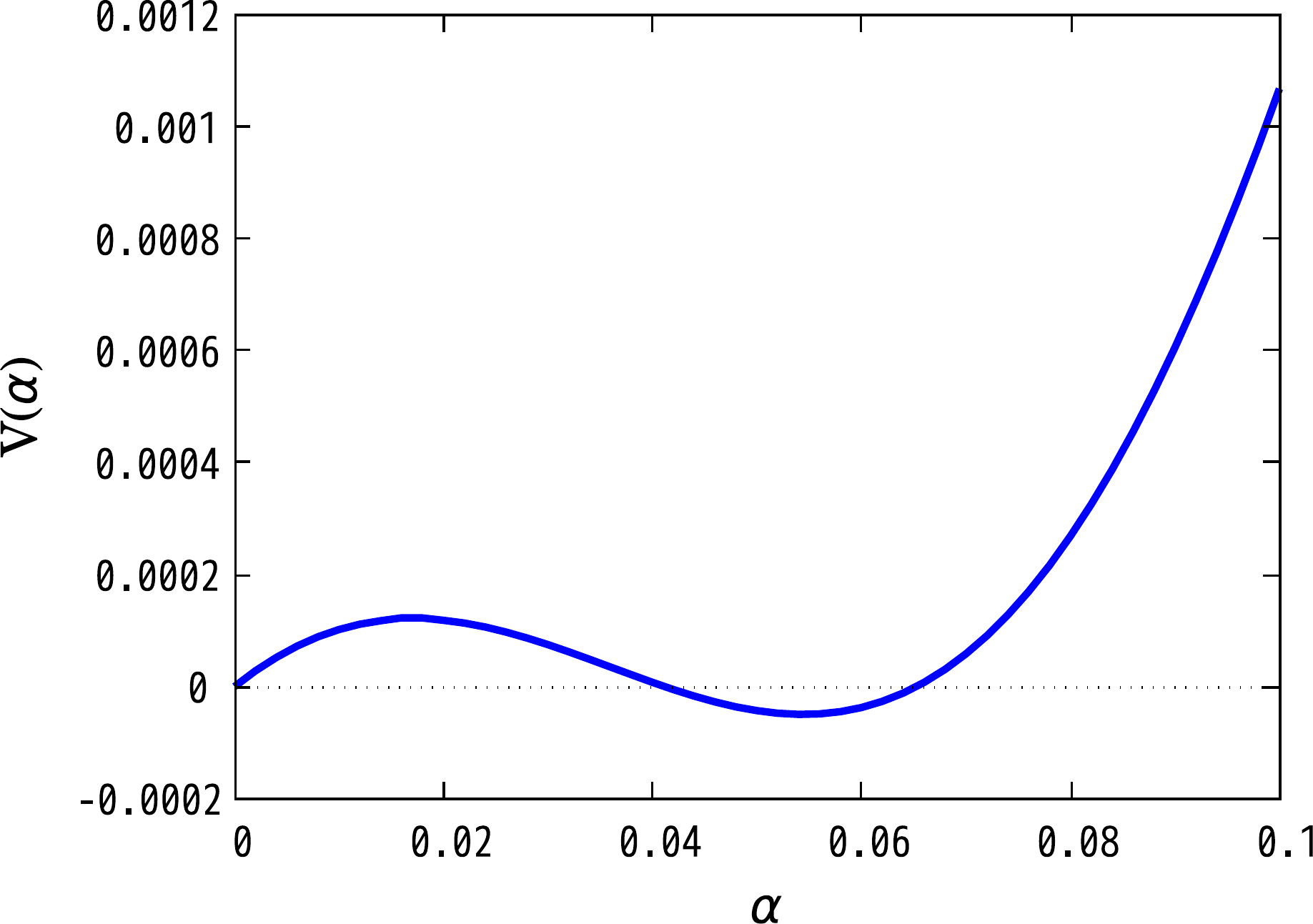}
\end{center}
\caption{
Higgs potential including 15 representation mirror fermions. The compactification scale $R^{-1} = 1.82{\rm TeV}$ and the bulk mass for the mirror fermion $M=0.8{\rm TeV}$ are chosen.
}
\label{fig:HP-all-realistic1}
\end{figure}

These observations indicate that the large contributions 
 from the extra fermions with the periodic boundary condition are necessary 
 for realizing a realistic electroweak symmetry breaking.
In this paper, we introduce the $\overline{\bf 15}$ representations as the mirror fermion
 because the period of the potential from the higher dimensional representations is smaller 
 and the curvature of the potential at the origin is more negative.  
Therefore, the potential is likely to realize the small VEV 
 as shown in the Figure \ref{fig:mirror}.
As shown in the Figure \ref{fig:HP-all-realistic1}, 
 the total effective potential of our model 
 has a minimum at the $\alpha = 0.0544$ 
 if we choose the compactification scale and the bulk mass for the third generation quarks 
 as $R^{-1}=1.82{\rm TeV},M=0.8{\rm TeV}$. 
In this case, 
 the Higgs boson mass $m_h=127{\rm GeV}$ and 
 the W boson mass $M_W=79.6\rm GeV$ are obtained.

\section{Summary}
In this paper, we proposed a new model of 5D $SU(3) \otimes U(1)_X$ GHU 
 with a successful electroweak symmetry breaking and a realistic Higgs boson mass.  
In our model, the representations of the fermions are very simple, 
 the $\bf 3, \bar{\bf 3}$ and $\overline{\bf 15}$ representations of $SU(3)$ gauge group.
Since top quark is not embedded into the $\overline{\bf 15}$ representations, 
 the anti-periodic boundary conditions can be imposed on $\overline{\bf 15}$. 
This reduced the number of the exotic massless fermions and the brane localized mass terms, 
 which largely simplifies our analysis. 

We have shown by calculating the 1-loop Higgs potential 
 that a realistic electroweak symmetry breaking and the observed Higgs mass are realized 
 in the case $R^{-1}=1.82 \text{TeV}, M=0.8\text{TeV}$. 
Note that a pair of additional $\overline{\bf 15}$ representations 
 other than the SM fermions have been introduced to accomplish the above result. 
The fact that the observed Higgs mass cannot be obtained 
 without extra fermions is consistent with the results 
 in the third and the fifth papers in \cite{Maru} and \cite{Maru:2017otg}. 
Furthermore, such extra fermions have been pointed out as the possible dark matter candidate \cite{Maru:2017otg}.   
We have also shown that the top and bottom quark masses are reproduced. 
As described in the main text, 
 this is not a trivial issue since these masses are correlated through the mixing 
 between the massless $SU(2)_L$ doublets from the up- and down-type sectors.

Finally, we give a comment on the relation to the SM-like property of the Higgs particle reported at LHC.   
Our Higgs potential has a periodicity with respect to the Higgs field 
because of the higher dimensional gauge symmetry.
It is significantly different from the SM,
however,
the small expectation values are required to happen the electroweak symmetry breaking 
and to obtain a realistic Higgs mass. 
In that case, the differences are generically small and consistent with the current experimental data. 


\section*{Acknowledgments}
The work of N.M. is supported in part by JSPS KAKENHI Grant Number JP17K05420.




\end{document}